\documentstyle[aps,multicol,psfig,epsf]{revtex} 
\begin{document} 
\topmargin -1.0truecm
\title{ \hfill
hep-ph/9906498\\ \vskip 1.5truecm
{\large{\bf Constraining Leptogenesis
from}}\\
{\large{\bf Laboratory Experiments}}} \vskip 2.0truecm

\author{ Anjan S. Joshipura$^{1,2}$ and Emmanuel A. Paschos$^2$\\
{\ns\it $^1$Theoretical Physics Group, Physical Research Laboratory,}\\
{\ns\it Navarangpura, Ahmedabad, 380 009, India.}\\
{\ns\it $^2$Inst. fur Theoretische Physik, Univ. of Dortmund,\\}
{\ns\it 44221, Dortmund, Germany}}
%\date{}
%------------------------------------------------------------
\def\ap#1#2#3{           {\it Ann. Phys. (NY) }{\bf #1} (19#2) #3}
\def\arnps#1#2#3{        {\it Ann. Rev. Nucl. Part. Sci. }{\bf #1} (19#2) #3}
\def\cnpp#1#2#3{        {\it Comm. Nucl. Part. Phys. }{\bf #1} (19#2) #3}
\def\apj#1#2#3{          {\it Astrophys. J. }{\bf #1} (19#2) #3}
\def\asr#1#2#3{          {\it Astrophys. Space Rev. }{\bf #1} (19#2) #3}
\def\ass#1#2#3{          {\it Astrophys. Space Sci. }{\bf #1} (19#2) #3}

\def\apjl#1#2#3{         {\it Astrophys. J. Lett. }{\bf #1} (19#2) #3}
\def\ass#1#2#3{          {\it Astrophys. Space Sci. }{\bf #1} (19#2) #3}
\def\jel#1#2#3{         {\it Journal Europhys. Lett. }{\bf #1} (19#2) #3}

\def\ib#1#2#3{           {\it ibid. }{\bf #1} (19#2) #3}
\def\nat#1#2#3{          {\it Nature }{\bf #1} (19#2) #3}
\def\nps#1#2#3{          {\it Nucl. Phys. B (Proc. Suppl.) }
                         {\bf #1} (19#2) #3} 
\def\np#1#2#3{           {\it Nucl. Phys. }{\bf #1} (19#2) #3}
\def\pl#1#2#3{           {\it Phys. Lett. }{\bf #1} (19#2) #3}
\def\pr#1#2#3{           {\it Phys. Rev. }{\bf #1} (19#2) #3}
\def\prep#1#2#3{         {\it Phys. Rep. }{\bf #1} (19#2) #3}
\def\prl#1#2#3{          {\it Phys. Rev. Lett. }{\bf #1} (19#2) #3}
\def\pw#1#2#3{          {\it Particle World }{\bf #1} (19#2) #3}
\def\ptp#1#2#3{          {\it Prog. Theor. Phys. }{\bf #1} (19#2) #3}
\def\jppnp#1#2#3{         {\it J. Prog. Part. Nucl. Phys. }{\bf #1} (19#2) #3}

\def\rpp#1#2#3{         {\it Rep. on Prog. in Phys. }{\bf #1} (19#2) #3}
\def\ptps#1#2#3{         {\it Prog. Theor. Phys. Suppl. }{\bf #1} (19#2) #3}
\def\rmp#1#2#3{          {\it Rev. Mod. Phys. }{\bf #1} (19#2) #3}
\def\zp#1#2#3{           {\it Zeit. fur Physik }{\bf #1} (19#2) #3}
\def\fp#1#2#3{           {\it Fortschr. Phys. }{\bf #1} (19#2) #3}
\def\Zp#1#2#3{           {\it Z. Physik }{\bf #1} (19#2) #3}
\def\Sci#1#2#3{          {\it Science }{\bf #1} (19#2) #3}
\def\n.c.#1#2#3{         {\it Nuovo Cim. }{\bf #1} (19#2) #3}
\def\r.n.c.#1#2#3{       {\it Riv. del Nuovo Cim. }{\bf #1} (19#2) #3}
\def\sjnp#1#2#3{         {\it Sov. J. Nucl. Phys. }{\bf #1} (19#2) #3}
\def\yf#1#2#3{           {\it Yad. Fiz. }{\bf #1} (19#2) #3}
\def\zetf#1#2#3{         {\it Z. Eksp. Teor. Fiz. }{\bf #1} (19#2) #3}
\def\zetfpr#1#2#3{         {\it Z. Eksp. Teor. Fiz. Pisma. Red. }{\bf #1} (19#2) #3}
\def\jetp#1#2#3{         {\it JETP }{\bf #1} (19#2) #3}
\def\mpl#1#2#3{          {\it Mod. Phys. Lett. }{\bf #1} (19#2) #3}
\def\ufn#1#2#3{          {\it Usp. Fiz. Naut. }{\bf #1} (19#2) #3}
\def\sp#1#2#3{           {\it Sov. Phys.-Usp.}{\bf #1} (19#2) #3}
\def\ppnp#1#2#3{           {\it Prog. Part. Nucl. Phys. }{\bf #1} (19#2) #3}
\def\cnpp#1#2#3{           {\it Comm. Nucl. Part. Phys. }{\bf #1} (19#2) #3}
\def\ijmp#1#2#3{           {\it Int. J. Mod. Phys. }{\bf #1} (19#2) #3}
\def\ic#1#2#3{           {\it Investigaci\'on y Ciencia }{\bf #1} (19#2) #3}
\def\tp{these proceedings}
\def\pc{private communication}
\def\ip{in preparation}
\relax

\newcommand{\GeV}{\,{\rm GeV}}
\newcommand{\MeV}{\,{\rm MeV}}
\newcommand{\keV}{\,{\rm keV}}
\newcommand{\eV}{\,{\rm eV}}
\newcommand{\Tr}{{\rm Tr}\!}
\renewcommand{\arraystretch}{1.2}
\newcommand{\beq}{\begin{equation}}
\newcommand{\eeq}{\end{equation}}
\newcommand{\beqa}{\begin{eqnarray}}
\newcommand{\eeqa}{\end{eqnarray}}
\newcommand{\ba}{\begin{array}}
\newcommand{\ea}{\end{array}}
\newcommand{\bmat}{\left(\ba}
\newcommand{\emat}{\ea\right)}
\newcommand{\refs}[1]{(\ref{#1})}
\newcommand{\ler}{\stackrel{\scriptstyle <}{\scriptstyle\sim}}
\newcommand{\ger}{\stackrel{\scriptstyle >}{\scriptstyle\sim}}
\newcommand{\lag}{\langle}
\newcommand{\rag}{\rangle}
\newcommand{\ns}{\normalsize}
\newcommand{\cm}{{\cal M}}
\newcommand{\gr}{m_{3/2}}
\newcommand{\p}{\partial}

\def\rp{ $R_P$} 
\def\321{$SU(3)\times SU(2)\times U(1)$}
\def\tl{{\tilde{l}}}
\def\tL{{\tilde{L}}}
\def\bd{{\overline{d}}}
\def\tL{{\tilde{L}}}
\def\a{\alpha}
\def\b{\beta}
\def\g{\gamma}
\def\c{\chi}
\def\d{\delta}
\def\D{\Delta}
\def\db{{\overline{\delta}}}
\def\Db{{\overline{\Delta}}}
\def\e{\epsilon}
\def\l{\lambda}
\def\n{\nu}
\def\m{\mu}
\def\nt{{\tilde{\nu}}}
\def\p{\phi}
\def\P{\Phi}
\def\x{\xi}
\def\r{\rho}
\def\s{\sigma}
\def\t{\tau}
\def\th{\theta}
\def\ne{\nu_e}
\def\nm{\nu_{\mu}}
\def\rp{$R_P$}
\def\mp{$M_P$}     
\renewcommand{\Huge}{\Large}
\renewcommand{\LARGE}{\Large}
\renewcommand{\Large}{\large}
\maketitle
\vskip 2.0truecm
\begin{center}
\underline{\bf{ABSTRACT}}
\end{center}
\vskip 1.0truecm
\begin{abstract}
The presently available information on neutrino oscillations
can be used to tightly constrain the light neutrino mass matrix $m_\n$
under the assumption that the three known neutrinos provide 
explanation of the solar and atmospheric anomalies. We exploit
left right symmetry and the seesaw model to establish direct correlation
between this constrained $m_\n$ and the mass matrix $M_R$ for the right handed
(RH) neutrinos. Using this correlation, one
could directly relate the lepton asymmetry
$\e$ generated in the decay of the lightest RH neutrino to experimentally
determined mixing angles of the light neutrinos. 
It is found that the 
parameters required to understand neutrino
anomalies also give rise to the $\e$ in the range required to reproduce
the correct baryon asymmetry in the leptogenesis scenario.
Specifically, the small angle MSW or the the vacuum oscillation 
explanation of the
solar anomaly can  lead to correct $\e$ for maximal value of the 
CP violating phases. In contrast, the large angle MSW solution needs
some suppression in CP violating phases in order to generate the right
$\e$. 
\end{abstract}
\newpage 
It is a well-known hypothesis \cite{fy} that the observed 
asymmetry in the Baryon number ($B$) of the universe owes its 
existence to lepton number ($L$) or lepton flavour violating \cite{smir}
interactions occurring at temperatures much greater than the weak scale.
According to this hypothesis-known as Leptogenesis-the  asymmetry in
Lepton number is created
by the out of 
equilibrium $L$ and CP violating interactions  at a high scale\cite{fy}. 
This can  get converted to the Baryon
asymmetry by the presence of non-perturbative
$B+L$ violating interactions \cite{sph} which are believed to be in
thermodynamic equilibrium
between the temperature $T\sim 10^2 -10^{12} \GeV$ . These
interactions
respect $B-L$ and wash out any asymmetry in $B+L$ leaving a residual
$B$ asymmetry $Y_B$\cite{kt}
\beq  Y_B= C Y_{B-L}\; , \eeq
C being a constant of order one determined by the content of the theory
at the electroweak scale.

The above scenario for Leptogenesis can be nicely realized in
the presence of heavy  right handed (RH) Majorana neutrinos\cite{fy,luty}.
The
existence of
these
neutrinos at a high scale is indirectly hinted by the observed (mass)$^2$
difference of the light left handed neutrinos at the Superkamioka \cite{sk}.
The scale $m_{\n_{\tau}}\sim 10^{-1}-10^{-2} \eV$ finds a natural
explanation in terms of the seesaw model with a right handed neutrino 
having mass near the grand unification scale. If the right
handed
neutrinos also display hierarchy similar to the left handed ones then
the decay of the lightest RH neutrino may be responsible for  generation
of the lepton asymmetry\cite{fy,luty}. The value of the $ Y_{B-L}$
generated
this way however depends upon the mixing and masses among the right handed
neutrinos. These are a priori arbitrary. In contrast, the masses
and mixing among
the light left handed neutrinos get strongly constrained  if
one demands
that the three light neutrinos simultaneously explain the observed 
solar and the  atmospheric neutrino deficits. If one could use the
underlying left right symmetry to relate the constrained spectrum of the
left handed neutrinos to that of the right handed ones, then  the
lepton asymmetry could be related to the low energy parameters. 
We wish to discuss here a possible scenario where such
relation comes out naturally. 

Let us consider conventional left right symmetric picture with three
right handed neutrinos $N_i\;,(i=1,2,3)$ mixing among each other. The  Dirac
Yukawa couplings of $N$ to the leptonic doublets $l$ leads to the decay
$N\rightarrow \phi\; l$ and
$N\rightarrow \phi^{\dagger}\; l^c$. The corresponding rates are
asymmetric in the presence of $CP$ violation. The lepton asymmetry
is generated  in this picture by the out of equilibrium decay of the
lightest right handed neutrino. The decay of this neutrino leads to
a CP   asymmetry $\epsilon$ \cite{luty}
\beq \label{eps}
\epsilon= {3\over 16\pi v^2}{1\over (m_D^{\dagger}m_D)_{11}}
\sum_{j=2,3} Im\left[ (m_D^{\dagger}m_D)_{1j}^2\right]{M_1\over
M_j}\; . \eeq

$m_D$ denotes the Dirac type
neutrino mass term and $M_i$ are the 
masses of the RH neutrinos. $v\sim 174 \GeV$ denotes the weak scale. 
We assumed hierarchical masses $M_i$ in writing the above $\epsilon$.
The $\epsilon$ gets resonantly enhanced if two of the RH neutrinos are
nearly degenerate \cite{pas}. We shall not consider this case and
concentrate on the above $\epsilon$.

The lepton asymmetry generated in the decay is related approximately to
$\epsilon$ \cite{luty} as follows:
\beq \label{db}
Y_L\sim \kappa {\epsilon \over g^*}\sim (10^{-3}-10^{-4})\;\; 
\epsilon
\; ,\eeq
where $g^*\sim 100$ are the effective degrees of freedom present 
in the cosmic plasma at $T\sim
M_1$. The $\kappa$ is a dilution factor arising due to 
other lepton number violating processes which tend to erase the lepton
asymmetry
in the decay. Typical value of $\kappa$ following from the numerical  
solution
of the relevant Boltzmann equations is around $10^{-1}-10^{-2}$
\cite{luty} leading to
the value
quoted on the RHS of eq. (\ref{db}). In view of the uncertainty in some of
the low energy parameters we shall not attempt to solve the exact Boltzmann
equation and determine $\kappa$. Instead, we shall take  $\kappa$ to be
in the range $10^{-1}-10^{-2}$ obtained in other analyses\cite{luty}. This
then requires through
eq.(\ref{db}) that one needs $\epsilon\sim (10^{-6}-10^{-7})$ in order to
generate
$Y_B\sim 10^{-10}$. Our main aim is to relate $\epsilon$ to masses and
mixing of the light neutrinos and see if one could get an $\epsilon$ in 
this range.

The CP asymmetry $\epsilon$ in the decay of a neutrino can  lead to
appreciable  $Y_L$  if it is out of equilibrium at the time of its decay.
This requires that 
\beq  \label{rate}
K\equiv {\Gamma_1\over H}={(m_D^{\dagger} m_D)_{11} M_1 \over 8 \pi
v^2H}\leq 1 \; ,\eeq
$H$ being the Hubble parameter. The above equation  places an important
constraint on the scale of the left right
symmetry breaking as we shall see.

Without assuming any specific model, we shall use the general structure
for
the neutrino masses that would follow in seesaw model with a left right
symmetry. The light neutrino masses in such scheme can be described by
\cite{pm}
\beq
\label{type2}
m_\n\sim m_{LL}- m_D M_R^{-1}m_D^T \; .\eeq
Here $m_{LL}\; (M_R)$ are Majorana masses for the left (right) handed
neutrinos and $m_D$ are Dirac masses connecting them. The assumption of
the left
right symmetry implies
\beq \label{lr}
m_{LL}\equiv v_L~f~~~~~~~~~~~~~~~~~~~M_R\equiv v_R~f \;\;,\eeq
where $v_R (v_L)$ are the (triplet) vacuum expectation values (vev)
which set the scale of the right (left) handed neutrino masses and $f$
is a Yukawa coupling matrix. One therefore has
\beq 
M_{R}={v_R\over v_L} m_{LL}\; . \eeq
 The vevs $v_{R,L}$
are not independent in a large class of models but are related by \cite{mg}
\beq 
\label{induced}
v_L\; v_R\sim \g M_W^2 \; , \eeq
$\g$ being a model-dependent parameter. Eq.(\ref{induced}) implies that
the
$v_L$ and hence the first term $m_{LL}$  in eq.(\ref{type2}) displays
seesaw
structure similar to the second term. $m_{LL}$ 
is normally presumed to be small and is neglected in conventional seesaw
model.  As emphasized
in \cite{cm}, $m_{LL}$ can  play an important
role in simultaneous solution of the solar and atmospheric neutrino
problems through seesaw mechanism. Eqs.
(\ref{type2},\ref{lr},\ref{induced}) are our
crucial
assumptions which would allow
us to fix the structure of $M_R$. 
We shall also assume that $m_D$ is determined by the up
quark masses and assume that the mixing among the up quarks and charged
leptons is small enough to be neglected.

We shall not presuppose any specific structure for the Yukawa coupling
matrix $f$ but determine it by requiring that the resulting $m_\n$
explains the solar and atmospheric anomalies simultaneously. The 
relevant scales can be identified as follows. The dominant
contribution of the seesaw term is of $O({m_t^2\over v_R})$. Requiring that
this is not larger than the atmospheric scale $\sim 10^{-1} \eV$ gives
$v_R \geq 10^{14} \GeV$, close to the grand unification scale.
Eq.(\ref{induced}) then
implies that $v_L$ is also close to the atmospheric neutrino scale. Thus
the contribution of the first term can be comparable to the largest
contribution from the conventional seesaw term and it is more appropriate
to retain both these terms. We shall do that in the following. 

Before proceeding further, let us collect relevant information on elements
of $m_\n$ following from experiments \cite{gu}. The present
experimental results
emerging from the data on atmospheric
neutrinos at Superkamioka and various solar neutrino detectors can be
reconciled in three neutrino framework either by assuming a
hierarchical or almost degenerate neutrino mass spectrum. 
We shall
concentrate in this paper on the hierarchical spectrum. The present
experimental results are strong enough to constrain elements of the
mixing matrix $U_L$.
The $U_L$ relates the flavour $\a=e,\m,\tau$
and mass $i=1,2,3$ eigenstates as follows:
\beq \nu_{\a}=U_{L\a i}\n_i \; .\eeq
After appropriate redefinition of the charged lepton states, the matrix
$U_L$
contains three mixing angles and three physical phases and can be
parameterized by \cite{param}
\beq
\label{u}
U_L=
W_{23}(\theta_{L23})\;W_{13}(\theta_{L13})\;W_{12}(\theta_{L12}e^{i\d})P(\l_i)\;,
\eeq
where $ P(\l_i)=diag. (e^{i\l_1},e^{i\l_2},1)$; $W_{ij}$ denote complex
rotation in the $ij$ plane. For example,
\beq
W_{12}(\theta_{L12} e^{i\d})\equiv\left(
\ba{ccc}
c_{L12}&s_{L12}e^{-i\d}&0\\
-s_{L12}e^{i\d}&c_{L12}&0\\
0&0&1\\
\ea \right).
\eeq
The $\delta$ in the above equation is analogous to the Kobayashi Maskawa
phase while $\l_{1,2}$ arise due to Majorana nature of neutrinos. The
latter cannot be probed through the conventional neutrino oscillation
experiments but
would play an important role in generating the lepton asymmetry in the
decay
of the right handed neutrinos as we shall see.

The constraints on different
elements in this case have been discussed in \cite{gu} and we
choose parameters according to values given in the first of \cite{gu}. The
simplifying
feature of this analyses is 
strong restriction implied by the CHOOZ experiment on the angle
$s_{L13}$\cite{gu}:
\beq \label{chooz} |U_{Le3}|^2\sim s_{L13}^2\leq 5\times 10^{-2} \; .
\eeq
This restriction is valid for $\Delta m_{31}^2\geq 3\cdot 10^{-3}\eV^2$.

Relative smallness of $s_{L13}$ together with hierarchy in masses
imply that the survival probabilities for the solar and atmospheric (ATM)
neutrino data assume two generation like form and restrict the
mixing angles $s_{L12},s_{L23}$ respectively. The  restrictions however depend upon the
chosen solution for the solar neutrino anomaly and there are three possibilities 
namely, small angle MSW \cite{msw} (SAMSW), large angle
MSW (LAMSW) and the vacuum oscillations (VAC). The relevant restrictions as presented for example in
\cite{gu} are  as follows:
\beqa \label{e1}
|U_{Le2}|&\approx s_{L12}&\approx 0.03-0.05\;,\;\;\;\;\;\;\;\;\;{\rm(SAMSW)}
\nonumber \\
&~~~~~~~~\approx &0.35-0.49 \;,\;\;\;\;\;\;\;\;\;\;\;\;\;{\rm (LAMSW})\nonumber \\
&~~~~~~~~\approx& 0.48-0.71 \;.\;\;\;\;\;\;\;\;\;\;\;\;\;{\rm (VAC)}\nonumber \\
|U_{L\m 3}|&\approx s_{L23}\approx& 0.49-0.71
\;.\;\;\;\;\;\;\;\;\;\;\;\;\;{\rm (ATM)} 
\eeqa
The two neutrino masses $m_{2,3}$ under the assumption of the hierarchy in
them
get constrained as follows:
\beqa \label{e2}
m_2^2&~~~\approx \Delta\;m_{SUN}^2\approx& (0.4- 1.2)\cdot 10^{-5} \eV^2 
\;,\;\;\;\;\;\;\;\;\;{\rm (SAMSW)}\nonumber \\
&~~~~~~~~~~~~~~~~~\approx& (0.8-3.0)\cdot 10^{-5} \eV^2 
\;,\;\;\;\;\;\;\;\;\;{\rm (LAMSW)} \nonumber \\
&~~~~~~~~~~~~~~~~~\approx&( 0.6-1.1)\cdot 10^{-10} \eV^2 
\;.\;\;\;\;\;\;\;{\rm (VAC)} \nonumber \\
m_3^2&\approx \Delta\;m_{ATM}^2~~\approx& (0.4-6.0)\cdot 10^{-3} \eV^2 
\;.\;\;\;\;\;\;\;\;\;{\rm (ATM)}
\eeqa
It is seen that apart from the phases, the mixing matrix $U_L$
is strongly restricted by the current experiments. This allows
reliable determination of the possible neutrino mass matrix and then of
$M_R$ through the left right symmetry.
In what follows we shall allow
$s_{L13}$ to vary between 0.0-0.22. However for analytic study we shall 
often neglect
$s_{L13}$  and concentrate on the following approximate  form of
$U_L$:
\beq
\label{app}
U_L\sim \left(
\ba{ccc}
c_{L12}e^{i \l_1}&s_{L12}e^{i (\l_2-\d)}&0\\
-s_{L12}c_{L23}e^{i(\d+\l_1)}&
c_{L12}\;c_{L23}e^{i\l_2}&s_{L23}\\
s_{L12}\;s_{L23}e^{i(\d+\l_1)}&
-c_{L12}\;s_{L23}e^{i\l_2}&c_{L23}\\
\ea \right). \eeq
Given the masses and mixing as in eq.(\ref{e1},\ref{e2}), we could
determine
$m_{\n}$ as
\beq
\label{mnu}
m_{\nu}=U_L^*\;diag.(m_1,m_2,m_3)U_L^{\dagger}\;\;.
\eeq
This in turn can be used to determine the Yukawa coupling matrix $f$
and hence $M_R$. Determination of $M_R$ in terms of the constrained low
energy parameters then directly leads to $\epsilon$.

It is seen from eqs.
(\ref{type2},\ref{lr},\ref{induced}) that ordinary seesaw contribution 
is suppressed
compared to $m_{LL}$ only
if the coefficient $\gamma$ is chosen much larger than 1. In the most
natural situation with $\gamma\sim O(1)$ both $m_{LL}$ and  
the top quark contribution to the second term of eq.(\ref{type2})
are comparable. 
We thus retain the seesaw contribution but we shall include only the
dominant top quark contribution. This contributes only
to $(m_\n)_{33}$ in eq.(\ref{type2}) when mixing among the up quarks is
neglected. Thus the Yukawa coupling matrix $f$
may be written as:
\beqa
\label{yuk}
f_{ij}&=&{(m_\n)_{ij}\over v_L},~~~~~~~~~{\rm when}~~{\rm both}~~i~~ {\rm
and}~~ j ~~{\rm are}
\neq 3 \nonumber \\
f_{33}&=&{(m_\n)_{33}+s\over v_L}.
\eeqa
The parameter $s$ refers to the contribution arising from the
seesaw term. This can be self consistently determined by inserting
eq.(\ref{yuk}) in eq.(\ref{type2}). One finds
\beq
\label{delta}
s\sim {v_L m_t\over \sqrt{\gamma}M_W}+O({v_L m_1\over m_3 s_{L12}^2}).
\eeq
As expected  $s$ is  comparable to $(m_\n)_{33}$ for
$\gamma\sim O(1)$ and $v_L\sim m_3$.
We have checked numerically that approximation made in eq.(\ref{yuk}) is
self-consistent, i.e. with $f$ determined from eqs.(\ref{yuk}), seesaw
contribution to elements of $m_{\n}$ other than the 33 is sub dominant for
$v_R\sim 10^{16} \GeV$ and $\gamma\sim O(1)$.

Eqs.(\ref{lr},\ref{yuk},\ref{delta}) together determine $M_R$ in terms
of the low
energy
parameters. This then allows us to obtain masses and mixing among the
right handed neutrinos. Let $U_R$ be defined as:
\beq
U_R^T M_R U_R=diag. (M_1,M_2,M_3) .\eeq
Parameterize $U_R$ as
\beq
\label{ur}
U_R=W_{23}(\theta_{R23} e^{i \phi_{R23}}) ~~W_{13}(\theta_{R13} e^{i
\phi_{R13}})~~~W_{12}(\theta_{R12} e^{i \phi_{R12}})~~ D_R.\eeq 
$M_R$ determined from eq.(\ref{type2}), eq.(\ref{lr}) and eq.(\ref{yuk})
can be diagonalized by above $U_R$ and its parameters can be determined
in terms of  elements of
$U_L$ and $s$:
\beqa \label{vacangle}
\tan 2 \theta_{R23}\approx {m_3 \over s}&~;~&\tan\phi_{R23}\approx
{s m_2 c_{L12}^2\sin 2 \l_2\over  m_3(m_3+s)}\;, \nonumber \\
\tan 2 \theta_{R13}\approx -\sin 2 \theta_{L12}{\sqrt{2}
m_2 c_{R23} Z_-\over
(m_3 Z_++2 s)}&~;~&\phi_{R13}\approx \d-2\l_2\;,\nonumber \\
\tan 2 \theta_{R12}\approx \sin 2 \theta_{L12}{\sqrt{2} m_2 c_{R23} 
Z_+\over
m_3 Z_-}&~;~&\phi_{R12}\approx \d-2\l_2\; .
\eeqa
\beq
D_R=diag.(e^{-i(\d-\l_2)},1,1)\;. \eeq
The masses of the RH neutrino are determined as:
\beqa \label {vacmasses}
M_1&\approx& {v_R\over v_L}  m_2 s_{L12}^{2}\; ,\nonumber \\
M_2&\approx& {v_R\over v_L}  {m_3 Z_-\over 2}\; ,  \nonumber \\
M_3&\approx& {v_R\over v_L} ({1\over 2} m_3 Z_+ +s). 
\eeqa
In the above equations,
$$ Z_{\pm}\equiv 1\pm{\sqrt{m_3^2+s^2}-s\over m_3} \; .$$
In deriving the above, we have assumed $m_1\ll m_2$ and specialized to the
$U_L$, eq.(\ref{app}) with maximal $s_{L23}$. Non-leading terms 
of $O({m_2\over m_3},{m_2\over s})$ are neglected above. As is evident, 
the mixing
pattern among RH neutrinos is completely determined  in terms of parameters of
$U_L$ and the seesaw contribution $s\sim m_3$. In the absence of the
seesaw contribution $s$, one would have got $U_R=U_L$ and $M_i={v_R\over 
v_L} m_i$. This is changed considerably by inclusion of $s$.
The mixing
angle $\theta_{R23}$  is large but not maximal. $\theta_{R12}$
is suppressed compared to  the corresponding
$\theta_{L12}$. A small $\theta_{R13}$ is induced although
corresponding $\theta_{L13}$ was chosen zero.  
The mass of the lightest neutrino is governed in this approximation by the
mass $m_2$ of the second generation neutrino rather than $m_1$ which is
neglected in the above derivation. The  RH neutrino
masses also display strong hierarchy in general. However it is seen from
eqs.(\ref{vacmasses}) that $M_2$ could become comparable to $M_1$
for some range of parameters in case of the large angle MSW solution to
the solar neutrino problem.

The out of equilibrium constraint on $N_{1}$ decay, eq.(\ref{rate}) now
implies
\beq \label{oed}
M_1\geq (2.8 \cdot 10^{16} \GeV) |U_{R31}|^2 .\eeq
Using $M_1$ and $U_{R31}$ from eq.(\ref{vacangle},\ref{vacmasses}), we
get,
 \beq
v_R\geq 2\;\; (2.8 \cdot 10^{16} \GeV) {v_L m_2\over m_3^2} c_{R23}^{2}
\left({s_{23}Z_+\over Z_-}-{c_{R23}Z_-\over Z_++2 s/m_3}\right)^2\; .
\eeq
We have retained only the leading contribution due to top quark in
writing eq.(\ref{oed}) and eq.(\ref{appeps}) below.
Since $v_L\sim m_3$ and factor in the last bracket is O(1) for $s\sim
m_3$ one
obtains
\beq
v_R\geq (2.8 \cdot10^{16} \GeV) {m_2\over m_3}\; . \eeq
The out of equilibrium condition is 
seen to be satisfied easily with $v_R \leq M_{GUT}$ independent of 
the chosen solution for the solar neutrino anomaly.

The $\e$ of eq.(\ref{eps}) is now given by
\beqa \label{appeps}
\e&\sim& {3\over 16 \pi}\left( {m_t\over v}\right )^2 \sin 2 \l_2
 \left[ c_{R12}^2s_{R23}^{2} {M_1\over M_2}+
 c_{R23}^{2} {M_1\over M_3}\right ]\; , \nonumber \\
&\sim&{3\over 8 \pi} \left( {m_t\over v}\right )^2\sin 2 \l_2
 \left[{m_2 s_{L12}^{2}\over m_3}\right]
\left[{c_{R12}^2s_{R23}^{2}\over Z_-} + {c_{R23}^{2}\over Z_++2s/m_3}
\right ]\; ,
\nonumber \\ 
&\sim&  10^{-1} \sin 2 \l_2 \;\left[{m_2 s_{L12}^{2}\over
m_3}\right] O(1)\; .\eeqa

This asymmetry crucially depends upon the phase $\l_2$ arising due to the
Majorana nature of neutrino. This phase cannot be constrained from the
oscillation data. Apart from this, $\epsilon$ is determined by the
known low energy parameters. The factor in the last bracket in
eq.(\ref{appeps})
is suppressed both for the MSW as well as the vacuum solution to the solar
neutrino problem. Typical value of the $\e$ in these cases is given by
\beqa
\epsilon&\sim& 5\times 10^{-6}\sin 2 \l_2\; ,\;\;\;\;\; {\rm (VAC)}
\nonumber \\
&\sim& 10^{-6}\sin 2 \l_2\; ,\;\;\;\;\;\;\;\;\;\;{\rm (SAMSW)}
\nonumber \\
&\sim&5\times 10^{-4}\sin 2 \l_2\; .\;\;\;\;\;\; {\rm (LAMSW)}
\eeqa
It is seen that except for the case of the large angle MSW solution one
gets the required $\epsilon$ for maximal value of the CP violating phase.

We have neglected contribution of the $s_{L13}$ in the above analytic
discussion. This angle is in fact allowed to be larger than the
$s_{L12}$
in case of the small angle MSW solution. Moreover, we have assumed
$m_3\sim s$ in the above  estimates. We now study
variation of $\epsilon$  numerically 
without making these approximations on $s$ and $s_{L13}$.

The input values for $s_{L23}$ and $m_3$ are fixed by the
atmospheric neutrino anomaly. The values for $s_{L12},m_2$
depend upon the chosen solution for the solar neutrino problem.
We consider all three possibility namely small angle MSW, large angle MSW 
and vacuum solutions. The variations of $|\epsilon|$ (eq.(\ref{eps})) 
and $K$ (eq.(\ref{rate})) with $s_{L13}$ and $s$ are displayed in
Figs. 1-3 respectively for  three different possibilities
mentioned above. These figures correspond to assuming large phases
namely $\delta=\pi/6,\l_1=\pi/3$ and $\l_2=\pi/4$. The $\epsilon$ is
largely insensitive to
chosen $\delta,\l_1$. 

Remarkably, for the experimentally determined parameters, one is able to
get  required $\epsilon$ and also satisfy out of equilibrium constraint
namely
$K\leq 1$. $\epsilon$ obtained in case of the SAMSW (Fig. 1) and VAC (Fig.
3) is in the correct range when  CP violating phases are chosen large as
in Figs.(1-3).
In contrast,
the LAMSW (Fig. 2) gives larger $\epsilon$ and thus would require
non-maximal phases if appropriate $Y_B$ is to be generated. This feature
is understood from the approximate expression, eq.(\ref{appeps}) which
shows
that $\epsilon$ is suppressed in case of VAC by smaller mass ratio.
The latter is larger for the MSW case but this is compensated by 
a smaller $s_{L12}$ in case of SAMSW. Similar compensation is not obtained for the 
LAMSW and one thus gets relatively larger $\epsilon$.

It is seen that $K$ and $\epsilon$ increase with
decrease in the ordinary seesaw contribution measured by $s$.
The  normal seesaw contribution $s$ plays  an important role in
keeping neutrinos out of equilibrium and in generating appreciable
$\epsilon$.
Numerical studies \cite{luty} have shown that one does not strictly need
$K\leq 1$ and appreciable $Y_L$ can be generated even when $K\sim O(10)$.
This happens in the figure typically 
for $s\geq 0.1 m_3$.
 We have plotted these figures
assuming $v_R=10^{16}\GeV$. Smaller value for $v_R$ will lead to larger
$K$ than the one displayed in the figures. 

There have been number of earlier papers \cite{earlier} which have tried
to
obtain
the structure of the RH masses from theoretical assumptions. These
works concentrated on definite textures for the  quark and lepton masses
and tried
to arrive at an $M_R$ which could explain neutrino anomalies. 
Our main emphasize here was the important contribution due to
left handed triplet Higgs which arise in any generic seesaw model
based on the left right symmetric theory. This symmetry  has in fact 
played a very important role in our analyses. Unlike the earlier works, 
we have been 
able to obtain expression (see, eq.(\ref{appeps})) for $\e$ directly in
terms of
the light neutrino masses
and mixing and a model dependent seesaw contribution $s\sim m_3$.

Let us now summarize the basic approach and results of the analyses
presented in this paper. The  anomalies observed in atmospheric
and the solar neutrinos find their natural explanation in terms of
the standard left right symmetric theory with three light and three
massive RH  neutrinos. The experimental observations constrain the
structure of the
neutrino mass matrix fairly strongly under theoretical assumptions of very
small mixing among charged leptons. We have shown here that it is possible
to relate the structure of the left handed neutrino mass matrix so
determined to that of
the right handed neutrino mass matrix in a class of models with left right
symmetry. This allows us to correlate   the lepton asymmetry
generated in the decay of the lightest RH neutrino to experimentally
inferred mixings among the light neutrino states. It is indeed gratifying
that this correlation is successful, i.e. the set of parameters needed to
explain neutrino anomalies can also
easily account for the lepton asymmetry of the required magnitude
without fine tuning in the relevant CP violating phase.

{\bf Acknowledgments}: One of us (ASJ) wishes to thank Alexander von
Humboldt foundation for support. We both wish to thank BMBF for support 
under contract 05 HT9PEA5. We thank W. Rodejohann for carefully
going
through the manuscript and for helpful suggestions. 

\newpage
\begin{center}
\begin{figure}[h]
\hspace*{0.5cm}
\epsfxsize 15 cm
\epsfysize 15 cm
\epsfbox[25 151 585 704]{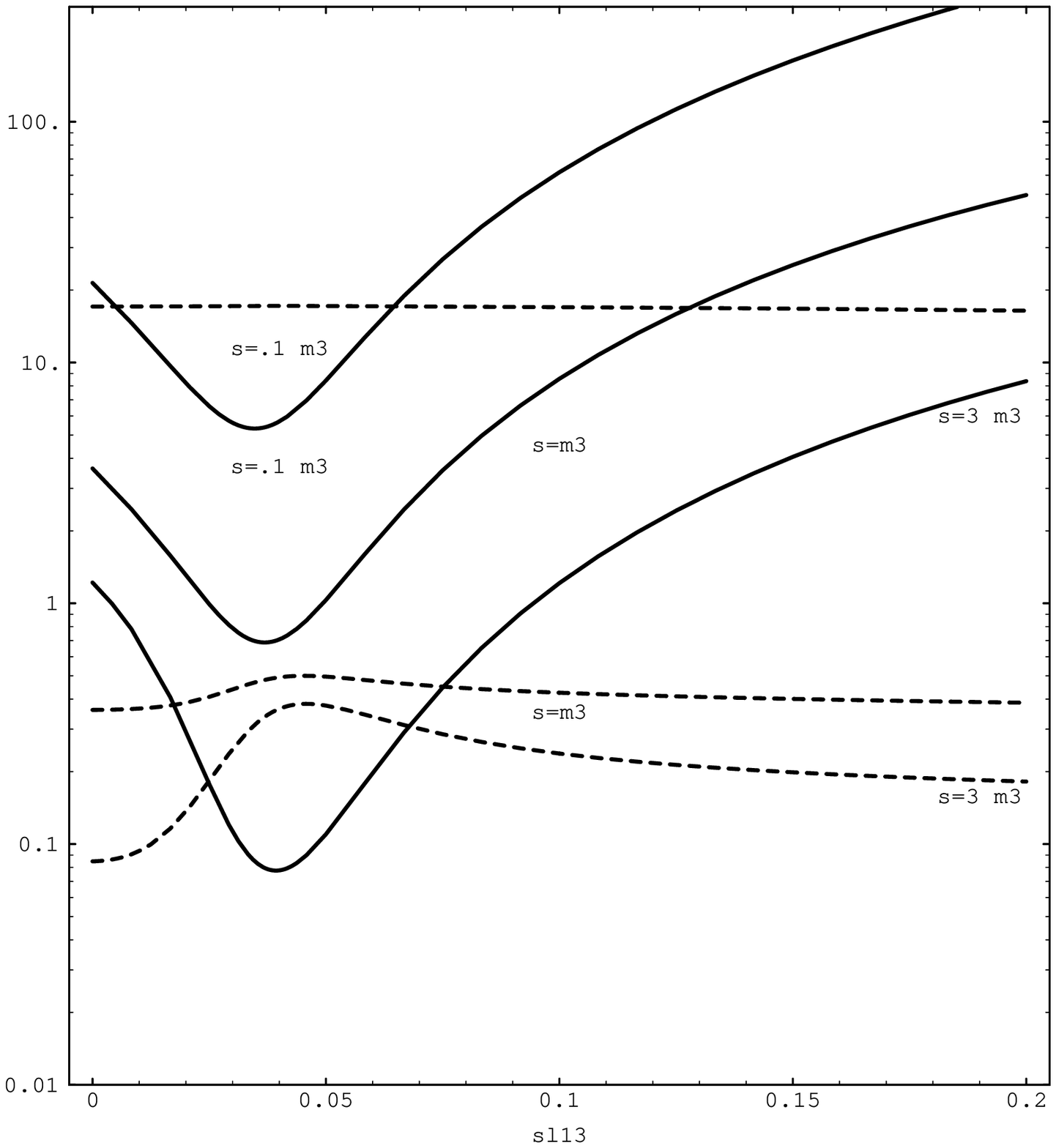}
\vspace*{.6cm}
\end{figure}
\end{center}
\noindent{\bf Figure 1}.~{\sl Variation of $10^6 |\epsilon|$ (solid line)
and $K$ (dotted line) with $s_{L13}$ for three different values of 
the seesaw contribution $s$ in case of the small angle MSW solution for
the 
solar neutrino problem. Values chosen for relevant parameters are
$s_{L23}=1/\sqrt{2},s_{L12}=.04,m_2=\sqrt{8\cdot
10^{-6}\eV^2},m_3=.07 \eV,\l_2=\pi/4,\delta=\pi/6,\l_1=\pi/3$.
\newpage
\begin{center}
\begin{figure}[h]
\hspace*{0.5cm}
\epsfxsize 15 cm
\epsfysize 15 cm
\epsfbox[25 151 585 704]{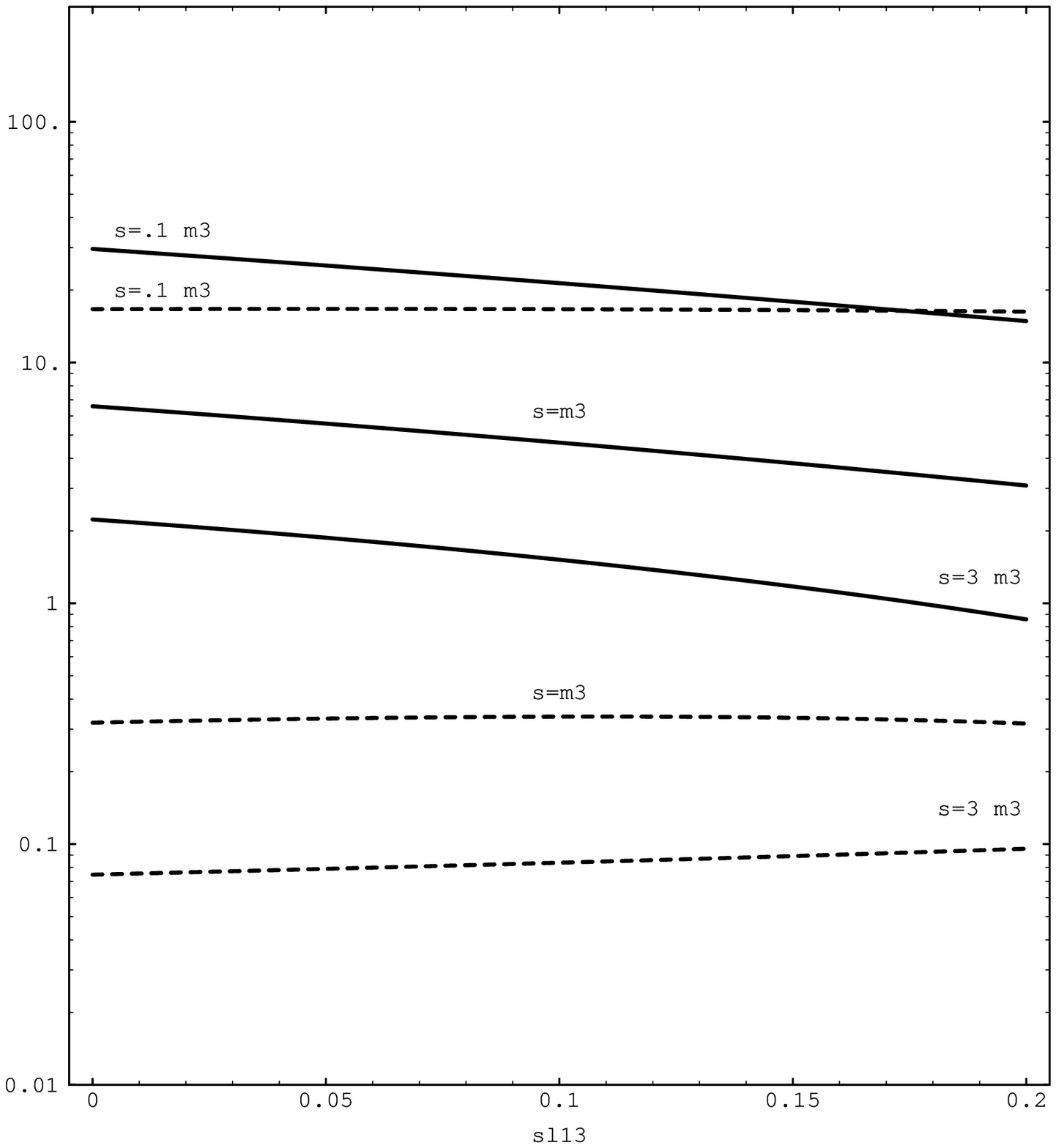}
\vspace*{.6cm}
\end{figure}
\end{center}
\noindent{\bf Figure 2}.~{\sl Variation of $10^4 |\epsilon|$ (solid line)
and $K$ (dotted line) with $s_{L13}$ for three different values of 
the seesaw contribution $s$ in case of the large angle MSW solution for
the 
solar neutrino problem. Values chosen for relevant parameters are
$s_{L23}=1/\sqrt{2},s_{L12}=.5,m_2=\sqrt{1.1\cdot
10^{-5}\eV^2},m_3=.07 \eV,\l_2=\pi/4,\delta=\pi/6,\l_1=\pi/3$.
\newpage
\begin{center}
\begin{figure}[h]
\hspace*{0.5cm}
\epsfxsize 15 cm
\epsfysize 15 cm
\epsfbox[25 151 585 704]{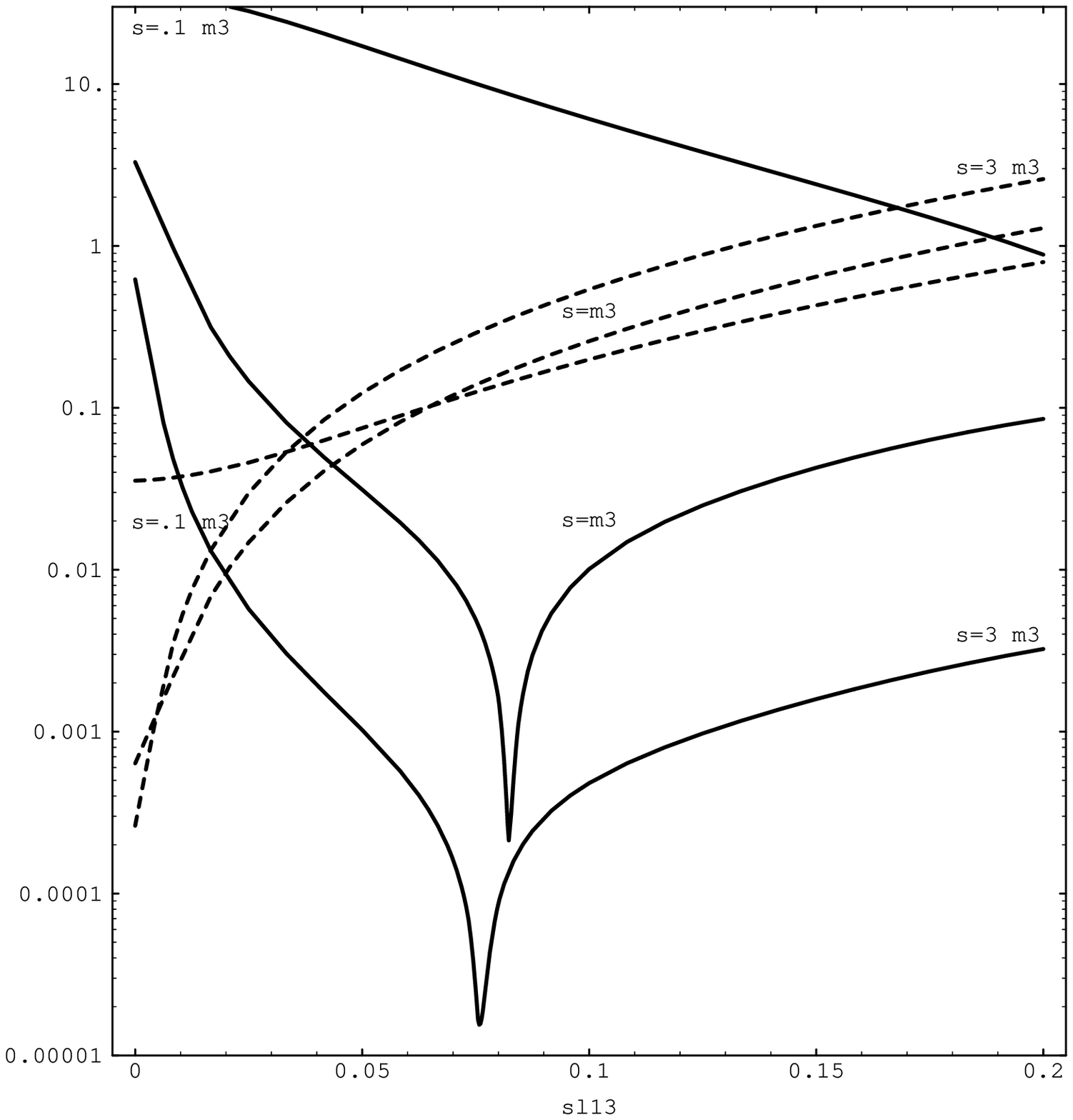}
\vspace*{.6cm}
\end{figure}
\end{center}
\noindent{\bf Figure 3}.~{\sl Variation of $10^6 |\epsilon|$ (solid line)
and $K$ (dotted line) with $s_{L13}$ for three different values of 
the seesaw contribution $s$ in case of the vacuum oscillation solution for
the 
solar neutrino problem. Values chosen for relevant parameters are
$s_{L23}=s_{L12}=1/\sqrt{2}, m_2=\sqrt{8\cdot
10^{-11}\eV^2},m_3=.07 \eV,\l_2=\pi/4,\delta=\pi/6,\l_1=\pi/3$
\end{document}